\documentclass[10pt, reqno]{amsart}

\usepackage{amssymb, amscd, mathrsfs, a4wide, color, cite, graphicx}

\newcommand{\hg}{{\hat g}}

\newcommand{\zJ}{\mathring{E}}

\newcommand{\bea}{\begin{eqnarray}}
\newcommand{\bean}{\begin{eqnarray}\nonumber}
\newcommand{\beal}[1]{\begin{eqnarray}\label{#1}}
\newcommand{\eeal}[1]{\end{eqnarray}\label{#1}}
\newcommand{\eea}{\end{eqnarray}}

\newcommand{\nohattau}{\tau}

\newcommand{\htau}{{\hat \tau}}

\newcommand{\mcO}{{\mycal O}}
\newcommand{\mcU}{{\mycal U}}

\newcommand{\mcNmM}{N^-_h\mcM}

\newcommand{\qedskip}{\hfill $\Box$\medskip}
\newcommand{\be}{\begin{equation}}
\newcommand{\eeq}{\end{equation}}
\newcommand{\beqa}{\begin{eqnarray}}
\newcommand{\eeqa}{\end{eqnarray}}
\newcommand{\beqan}{\begin{eqnarray*}}
\newcommand{\eeqan}{\end{eqnarray*}}
\newcommand{\ba}{\begin{array}}
\newcommand{\ea}{\end{array}}
\newcommand{\eq}[1]{(\ref{#1})}

\newcommand{\mcC}{{\mycal C}}

\newcommand{\beaa}{\begin{eqnarray*}}
\newcommand{\eeaa}{\end{eqnarray*}}

\numberwithin{equation}{section}

\newcommand{\dLp}{d_L^-}

\newcommand{\tf}{\tau_\varphi^+{}}
\newcommand{\tp}{\tau_\varphi^-{}}
\newcommand{\tpi}{\tau_{\varphi_i}^-{}}

\newcommand{\tpm}{\tau_\varphi^\pm{}}

\DeclareFontFamily{OT1}{rsfs}{}
\DeclareFontShape{OT1}{rsfs}{m}{n}{ <-7> rsfs5 <7-10> rsfs7 <10-> rsfs10}{}
\DeclareMathAlphabet{\mycal}{OT1}{rsfs}{m}{n}

\newcommand{\mcM}{{\mycal M}}

\theoremstyle{plain}
\newtheorem{theorem}{Theorem}[section]
\newtheorem{Theorem}[theorem]{\sc Theorem}
\newtheorem{corollary}[theorem]{\sc Corollary}
\newtheorem{proposition}[theorem]{\sc Proposition}
\newtheorem{lemma}[theorem]{\sc Lemma}

\theoremstyle{definition}

\theoremstyle{remark}

\newtheorem{remark}[theorem]{\sc Remark}

\newcommand\bel[1]{\begin{equation}\label{#1}}
\newcommand\ee{\end{equation}}

\newcommand\N{\ensuremath{\mathbb{N}}}
\newcommand\R{\ensuremath{\mathbb{R}}}

\newcommand\vs{\vskip .2cm}

\newcounter{mnotecount}[section]

\renewcommand{\themnotecount}{\thesection.\arabic{mnotecount}}

\newcommand{\mnote}[1]
{\protect{\stepcounter{mnotecount}}$^{\mbox{\footnotesize
$
\bullet$\themnotecount}}$ \marginpar{
\raggedright\tiny\em
$\!\!\!\!\!\!\,\bullet$\themnotecount: #1} }

\title{On differentiability of volume time functions}

\begin{document}
\author{Piotr~T.\ Chru\'{s}ciel}
\email{piotr.chrusciel@univie.ac.at}
\address{Fakult{\"a}t f{\"u}r Physik und Erwin Schr\"odinger Institut \\ Universit{\"a}t Wien \\ Boltzmanngasse 5 \\ 1090 Wien, Austria}

\author{James~D.E.\ Grant}
\email{j.grant@surrey.ac.uk}
\address{Department of Mathematics \\ Faculty of Engineering and Physical Sciences \\ University of Surrey \\ Guildford \\ GU2 7XH \\ U.K.}

\author{Ettore~Minguzzi}
\email{ettore.minguzzi@unifi.it}

\address{
Dipartimento di Matematica e Informatica ``U. Dini''
\\
Universit\`{a} degli Studi di Firenze\\ Via S. Marta 3\\ I-50139
Firenze, Italy}

\date{\today\protect\thanks{Preprint UWThPh-2013-1}}

\begin{abstract}
We show differentiability of a class of Geroch's volume functions on
globally hyperbolic manifolds. Furthermore, we prove that every
volume function satisfies a local anti-Lipschitz condition over
causal curves, and that  locally Lipschitz time functions which are
locally anti-Lipschitz  can be uniformly approximated by smooth time
functions with timelike gradient. Finally, we prove that in stably
causal spacetimes Hawking's time function can be uniformly
approximated by smooth time functions with timelike gradient.
\end{abstract}

\maketitle
\thispagestyle{empty}

\section{Introduction}

In his classical work on domains of dependence and global
hyperbolicity~\cite{GerochDoD} Geroch showed how to construct a time
function, namely a continuous function increasing over every future
directed causal curve, by considering a weighted
volume of the chronological past of the point. This strategy was
extended by Hawking~\cite{HawkingPRSL68,HE} who proved that in a
stably causal spacetime it is possible to obtain a
time function through suitable averages of Geroch's volume
functions
(see~\cite{Seifert,FathiSiconolfiTime,BernalSanchez2,MinguzziTime}
and references therein for alternative constructions). Nevertheless,
even in the globally hyperbolic case the differentiability
properties of the resulting functions do not seem to have been
properly understood so far.%
\footnote{The reader is referred to~\cite{sanchez05b,BernalSanchez2} for a review of the history of the problem.}
The object of this note is to establish
differentiability of a large class of volume functions under the
hypothesis of global hyperbolicity, as well as smoothability of a
class of time functions for stably causal space-times. Indeed,
assuming global hyperbolicity, at the end of
Section~\ref{s18XI12.11} below we prove:

\begin{Theorem}
 \label{T15XI12.1}
Let $(\mcM,g)$ be a globally hyperbolic spacetime with a $C^{2,1}$
metric. There exists a
class of smooth functions $\varphi>0$ such that the functions
\bel{17XI12.1}
 \tpm(p) = \int_{J^{\pm}(p)} \varphi  \, d\mu_g
 \;,
\ee
where $d\mu_g$ is the volume element of $g$, are
continuously differentiable with
timelike
gradient.
\end{Theorem}

Following Geroch, we can now define $\tau_\varphi=\ln
(\tau^-_\varphi/\tau^+_\varphi)$ so that if $\gamma$ is an
inextendible causal curve then $\tau_\varphi \circ \gamma$ is onto
$\mathbb{R}$. As a consequence, the differentiability
of $\tau_\varphi$ implies that the level sets of $\tau_\varphi$ are Cauchy
spacelike hypersurfaces (thus not just acausal and
Lipschitz).

In view of the analysis in~\cite{itoh00} it is conceivable that the
functions $\tpm$ are $C^{1,1}$,   but we have not
investigated the issue any further, as we will smooth out these
functions in any case, see Corollary~\ref{c27XII12.1} below. The smoothing procedure will establish the
existence of smooth Cauchy time functions in globally hyperbolic
spacetimes. This result, already obtained in~\cite{BernalSanchez,BernalSanchez1,FathiSiconolfiTime}
by different means, plays a key role in the theory
as it implies that
Geroch's topological splitting~\cite{HE} can actually be chosen
smooth.

Some comments on the proof might be in order. We write a light-cone
integral formula for a candidate derivative of $\tpm$. The
integrand involves Jacobi fields which might be blowing up as one
approaches the end of the interval of existence of the geodesic
generators of
$$
 E^\pm(p):=  J^\pm (p)
 \setminus I^\pm(p)
 \;.
$$
So the weighting function $\varphi$ has to compensate for this,
which provides one of the constraints on the set of admissible
functions $\varphi$. In particular, we are going to introduce an
auxiliary complete Riemannian metric $h$ in order to control and
obtain a sufficiently fast fall-off of $\varphi$ at infinity.

An interesting feature of our candidate formula for the derivative
of $\tpm$ is that it involves an integral on just $E^{\pm}(p)$, and
not on the whole light cone issued from $p$. As a consequence, most
of the pathological behavior connected with  non-differentiability
of the exponential map after conjugate points is avoided.

\begin{figure}
\begin{center}
{\includegraphics[scale=.7]{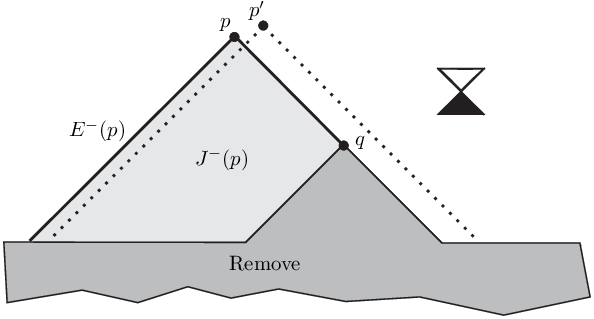}}
\caption{The volume of past light-cones is Lipschitz-continuous but not differentiable at $\dot J^+(q)$.%
\label{F24XI12.1}}
\end{center}
\end{figure}

Now, the
domain of integration that interests us is generated by lightlike
geodesics that  may be either complete or incomplete. The
distinction between completeness and incompleteness is, however,
rather unimportant since, without loss of generality, we may
conformally rescale $g$ so as to make all the null geodesics
complete~\cite{Beem76}.
Next, there might exist generators that do not meet the cut-locus
and which span an area region on $E^{\pm}(p)$ that does \emph{not}
vary continuously with $p$. (Example:
 Let $(N,g) $ be any bounded globally hyperbolic subset of two-dimensional Minkowski spacetime, with the weighting function $\varphi$ in \eq{17XI12.1} equal to one. Let $q\in N$ and set $M=N\setminus J^-(q)$, with the induced metric.  Then $\tp$ is not differentiable at the boundary of the future of $q$; see Figure~\ref{F24XI12.1}.)
 It turns out that the discontinuous behavior illustrated by the example happens
``close to infinity" in the complete auxiliary metric, where a fast fall-off of $\varphi$ amends the problem.

In Section~\ref{s18XI12.1} we show that a local application of our
integral formula gives a simple proof of the anti-Lipschitz
character of $\tpm$ along causal curves. We also show that the
anti-Lipschitz condition with respect to a spacetime metric with
wider light cones allows us to smooth the time function. This result
is used to prove the smoothability of Hawking's time function in
stably causal spacetimes, and to prove the existence of smooth
Cauchy time functions in globally hyperbolic spacetimes, by taking
advantage of the stability of global hyperbolicity. We also prove
the equivalence between stable causality and the existence of a time
function by taking advantage of the equivalence between the former
property and $K$-causality.

Finally, in the last section we prove that we can dispense with the
result on the stability of global hyperbolicity, and prove directly
the smoothability of Geroch's Cauchy time functions, by showing that
every Lipschitz and anti-Lipschitz time function is in fact
anti-Lipschitz with respect to a spacetime metric with wider light
cones.

\section{The null cut-locus} \label{hgc}

To avoid ambiguities, we start by noting that our signature is
$(-,+\cdots+)$, and that we use a convention in which the zero
vector is \emph{not\/} a null vector.
Space-times of any dimension $n+1$, $n\ge 1$, are allowed, though it must be said that the case $n=1$ is rather simpler than the remaining dimensions, as there are then no null conjugate points.

The proof of Theorem~\ref{T15XI12.1} will require some understanding
of the null cut-locus. For this, we start by recalling some
definitions and results from~\cite{BeemEhrlichEasley}, see
especially Sections~9.2 and 10.3 there.

Our spacetime metric $g$ will be $C^{2,1}$ throughout the paper. This
condition assures the existence of convex neighborhoods,
continuous differentiability of the exponential map and Lipschitzness of the Riemann tensor.  These conditions can possibly be weakened.

The \emph{past Lorentzian distance function}
$$\dLp \colon \mcM\times\mcM\to [0,\infty]
$$
is defined as
 \bel{15XI12.2}
  \dLp(p,q)=
  \left\{
    \begin{array}{ll}
      0, & \hbox{$q\not\in J^-(p)$;} \\
      \sup{\int_\gamma} \sqrt{-g(\dot \gamma,\dot \gamma)} , & \hbox{otherwise,}
    \end{array}
  \right.
\ee
where the $\sup$ is taken over all past-directed causal curves from
$p$ to $q$. 
In general smooth spacetimes the function $\dLp$ is lower
semi-continuous~\cite{BeemEhrlichEasley}, while for globally
hyperbolic (smooth) spacetimes it is finite and continuous. Moreover, we observe that
the arguments in~\cite{ChCausality,BeemEhrlichEasley} can be used to
prove these results for $C^2$ metrics. (For more
on the continuity properties of the Lorentzian distance function
under low causality conditions see~\cite{minguzzi08e}.)

Throughout this paper we choose once and for all a smooth complete
Riemannian metric $h$ on $\mcM$. We denote by $\mcNmM$ the bundle of
past-directed $h$-unit $g$-null vectors. A curve will be said to be \emph{$h$-parametrised} if it is parametrised with arc-length measured with the metric $h$.

Sometimes, for simplicity, we shall speak of {\em geodesics} though strictly speaking
we should speak of {\em pregeodesics}, namely when the curve
is a geodesic up to the parametrization.
We will
say that $\gamma$ is a \emph{half-geodesic\/} if $\gamma$ satisfies
the geodesic equation and if $\gamma \colon [0, a) \to \mcM$ is
maximally extended in the direction of increasing parameter. If
$\gamma$ is parametrised by $h$-arc length with respect to a
complete Riemannian metric, then $a=\infty$ (see,
e.g.,~\cite{ChCausality,MinguzziLimit,BeemEhrlichEasley}).

For any past-directed null half-geodesic $\gamma \colon [0,\infty)\to \mcM$  parametrised by $h$-arc length we set
\begin{eqnarray}
 t_-(\gamma) &= &\sup\big\{t\in \R^+: \dLp(\gamma(0),\gamma(t))=0
 \big\} \in [0,\infty]
 \;.
\label{15XI12.1}
\end{eqnarray}
The points of $E^{-}(p)$ which are sufficiently close to $p$ are connected to $p$ by
achronal geodesics starting at $p$. We now define a subset $\zJ^-(p)$ through a union of lightlike geodesic segments as follows
\begin{eqnarray}
 \zJ^-(p)
 &= &
\big\{ \gamma(s) \,| \ \gamma(0)=p\;,
 \  s\in(0,t_-(\gamma))
\big\}
 \;.
\label{18XI12.2}\end{eqnarray}
If $(\mcM,g)$ is globally hyperbolic and if $t_-(\gamma)<\infty$, then the point $\gamma(t_-(\gamma))$ is either
conjugate  to $\gamma(0)$ along $\gamma$ and/or there exist two
distinct null achronal geodesics from $\gamma(0)$ to
$\gamma(t_-(\gamma))$ (see~\cite[Theorem~9.15]{BeemEhrlichEasley},
 compare the arguments in the proof
of Proposition~\ref{P18XI12.1}).
 The
points $\gamma(t_-(\gamma))$ are \emph{end points\/} of generators
of past light-cones~\cite{BK2}. The set of end points of
past-directed generators starting at $p$ is called
\emph{the past null cut-locus} of $p$. It is known, in spacetime dimension $n+1$,
that the past null cut-locus of $p$ has vanishing $n$-dimensional measure within $\exp_p (\mathbb{R}(N_h^-)_p\mcM)$.
(This fact also follows from Fubini's theorem and Proposition~\ref{P18XI12.1} below; compare the proof of Lemma~\ref{L18X14.1}.)
  (The result is of course trivial in $1+1$ dimensions.)

For a $C^{2,1}$ metric $g$, the set $\zJ^-(p)$ is a $C^{1,1}$ null hypersurface, indeed it is an immersion by the local injectivity of the exponential map away from conjugate points, and it is an embedding because there are no self intersections as we remove the points of the light cones behind the cut points and the cut points themselves.  In a globally hyperbolic spacetime
\bel{18XI12.1}
 E^-(p)= \overline{ \zJ^- (p)}
 \;.
\ee

We can parameterize the set of all maximally extended past-directed half-geodesics
by
the initial positions and $h$-normalised tangent vector at the starting point. In other words, such half-geodesics are in one-to-one correspondence with vectors in $\mcNmM$. This induces in the obvious way a topology on the set of half-geodesics. We have:

\begin{proposition}
 \label{P18XI12.1}
  Let $(\mcM,g)$ be globally hyperbolic with a twice-differentiable metric.
Then the map $t_- \colon \mcNmM \to [0, \infty]$ is continuous.
\end{proposition}

{\noindent \sc Proof:}
The proof is adapted from that of the corresponding result in
Riemannian geometry given in~\cite[Prop.~5.4]{CheegerEbinSecondEd}, compare
\cite[Vol.\ II, p.\ 99]{kobayashi:nomizu}. Let $\gamma_i \colon [0,
\infty) \to \mcM$ be a sequence of past-directed null half-geodesics
parametrised by $h$-arc length  such that $\gamma_i(0) \to \gamma(0)$
and $\dot{\gamma}_i(0) \to \dot{\gamma}(0)$ as $i \to \infty$. By
continuous dependence on initial conditions of solutions of the
geodesic equation it holds that $\gamma_i(t) \to \gamma(t)$ as $i
\to \infty$ for each $t > 0$. We split the rest of the proof into
two steps:

\vs \noindent{1). {\sc Upper semi-continuity of $t_-$}}:
Let $t > 0$ and assume that there exists infinitely many $i$ such that
$\dLp(\gamma_i(0), \gamma_i(t)) = 0$. Since $\gamma_i(t) \to
\gamma(t)$ as $i \to \infty$, and $\dLp \colon \mcM \times \mcM \to
[0, \infty]$ is continuous, it follows that $\dLp(\gamma(0),
\gamma(t)) = 0$. Hence $t_-(\gamma) \ge t$. Therefore, $t_-(\gamma)
\ge \limsup_{i \to \infty} t_-(\gamma_i)$, as required.
(Note that there was nothing to prove when $t_-(\gamma)=\infty$.)

\vs \noindent{2). {\sc Lower semi-continuity of $t_-$}}:
We need
to show that $t_-(\gamma)\le \liminf_{i \to \infty} t_-(\gamma_i)$.
We assume that $\liminf_{i \to \infty} t_-(\gamma_i) < + \infty$,
otherwise there is nothing to prove. Let
$$
 t > \liminf_{i \to \infty}
t_-(\gamma_i) =:\ell
 \;,
$$
then there exists a subsequence, also denoted
$\{\gamma_i\}$, such that $t_-(\gamma_i) < t$ for all $i$ with
$t_-(\gamma_i) \to \ell< t$ as $i \to \infty$. Since $t_-(\gamma_i) < t$, we deduce that
$\gamma_i(t) \in I^-(\gamma_i(0))$, and therefore there exist
\emph{timelike} half-geodesics $\sigma_i$ parametrised by
$h$-arc length  such that $\gamma_i(0) = \sigma_i(0)$ and $\gamma_i(t) =
\sigma_i(\widetilde{t}_i)$ for some $\widetilde t_i \in
(0,\infty)$.

Passing to another subsequence if needed, by global hyperbolicity
there exists a causal past-directed half-geodesic $\sigma$ such that
$\sigma_i(s)\to \sigma(s)$ for $s\in [0,\infty)$, with the sequence
$\{\widetilde t_i\}_{i\in\N}$ convergent, and
$\sigma_i(\widetilde{t}_i)\to \gamma(t)$. If $\sigma$ and $\gamma$
are distinct, it follows that $t\ge t_-(\gamma)$, as desired. If $\sigma$
and $\gamma$ coincide but $t$ is larger than or equal to the
distance to the first conjugate point of $\gamma(0)$ along $\gamma$
we again obtain $t\ge t_-(\gamma)$, and we are done.

It remains to consider the possibility that $\sigma$ and $\gamma$
coincide and that $t$ is smaller than the $h$-distance (possibly infinite, if $t_-(\gamma)=\infty$)
to the first  conjugate  point along $\gamma$ of $\gamma(0)$. For $X\in T\mcM $ let $\gamma_X$
denote a half-geodesic such that $\dot \gamma_X(0)=X$. By
continuity of $\det \exp_*$, there exists a neighborhood $\mcO$ of $\dot \gamma(0)$
in $T\mcM$ such that for every causal vector $X\in \mcO$ and every
past-directed half-geodesic $s\mapsto \gamma_X(s)$ the $h$-distance along $\gamma_X$
to the first conjugate  point of $\gamma_X(0)$ is
larger than $t$.  This contradicts $t_-(\gamma_i)<t$ thus this case does not really apply.
Hence, $t_-$ is lower semi-continuous, and thus also continuous, as desired.
\qed

\section{The derivative of $\tpm$}
 \label{s18XI12.11}

In this section we assume that $(\mcM,g)$ is a globally hyperbolic spacetime and
show that the functions $\tpm$, as defined in equation~\eqref{17XI12.1},
are differentiable for suitably chosen $\varphi$.
We consider only $\tp$, the result for $\tf$ follows by changing time-orientation.
We will always assume that $\varphi$ is continuous and non-negative.
We start by assuming that $\varphi$ has compact support.

Let
$ p\in\mcM,
$
and let
\bel{8XI14.5}
 \gamma \colon \R\to\mcM
\ee
be any future-directed, timelike $h$-arc length parametrised curve passing through $p=\gamma(0)$.
Choose a $g$-orthonormal frame at $p$, and parallel-propagate the frame along $\gamma$.
This defines $g$-orthonormal frames $\{e_\mu(s)\}$ at $\gamma(s)$.
We will say that $g$-geodesics at different points of $\gamma$ are \emph{pointing in the same direction\/}
if the frame components of their initial velocities in the frame $\{e_\mu (s)\}$ coincide,
i.e. if their tangent vectors at $\gamma$ are parallel transports of each other along $\gamma$.
Then, for each generator of $\zJ^-(p)$, we may associate a family of half-geodesics, parametrised by $s \in \R$,
that emanate from the point $\gamma(s)$ with initial tangent vector pointing in the same direction as the chosen generator.
Thus, points on neighbouring light-cones with vertices on $\gamma$
can be obtained by flowing along the associated Jacobi fields.
This explains the construction that follows.

Let $\tau \mapsto \Gamma_s(\tau)$ be any past-directed affinely parametrised null half-geodesic starting at $\gamma(s)$,
where $\tau \in [0, \tau_-(s))$ , with $\Gamma_s(\tau_-(s))$ the cut point of $\Gamma_s$.
Its tangent vector $\frac{d}{ d \tau}$ at $\gamma(s)$ is extended all over $\gamma$ through parallel transport, i.e.
$$
 \frac{D}{d s} \frac{d}{ d \tau}=0 ,
$$
over $\gamma$. Taking this parallel tangent vector field as initial data in the geodesic equations,
the definition of $\Gamma_s$ is extended to different values of $s$. 
It is well known that by the local injectivity of the exponential map away from conjugate points,
$\tau_-(s)$ is lower semi-continuous, thus the pairs $(s, \tau)$, $s \ne 0$, for which $\Gamma$ is defined form an open set.
The mapping $\Gamma(s,\tau)=\Gamma_s(\tau)$ is generated by past-directed lightlike half-geodesics with initial endpoint $\gamma(s)$ at $\gamma$ and is really an embedding (surface). Indeed, two geodesics generators relative to different values of $s$ cannot intersect, namely it cannot be $\Gamma_{s'}(\tau')=\Gamma_s(\tau)$ for $s'>s$, otherwise it would be possible to go from $\Gamma_{s'}(\tau')$ to $\gamma(s')$ with a timelike curve in contradiction with the fact that $\Gamma_{s'}(\tau')$ stays before the cut point of $\Gamma_{s'}$.
Similarly, the image of $\Gamma$ cannot develop focusing points, for
this would imply that a certain Jacobi field $X$ to be introduced in a moment vanishes, a fact which we prove to be impossible.

Since $(s,\tau)$ provide coordinates over the image   of $\Gamma$, we have that

$$
 T:=\Gamma_*\partial_\tau \ \mbox{and} \  X:=\Gamma_* \partial_s
$$
commute near $\gamma$, thus
$$
 \frac{D}{d \tau} \frac{d}{ d s}=0
$$
over $\gamma$. Now, observe that $X$ is the variational field of $\Gamma_s(\tau)$, where the longitudinal curves are geodesics. Thus it is a Jacobi field whose value at $\gamma(s)$ is $\dot \gamma(s)$, while its first covariant derivative vanishes thanks to the mentioned commutation relation. It is interesting to observe that, since $X$ is Jacobi, for every fixed $s$ we have that $\tau\mapsto g(X,\frac{d}{d \tau})$ is an affine function of $\tau$ over $\Gamma_s(\tau)$, a fact which, given the initial conditions, implies that $g(X,\frac{d}{d\tau})$ is a constant whose value can be inferred from its value at the tip $\gamma(s)$. In particular, since $\dot{\gamma}$ is future-directed and timelike, $g(X,\frac{d}{d \tau})$ is positive over $\Gamma_s$ so, as anticipated, there cannot be focusing points due to the variation of coordinate $s$, although, of course, each individual light cone for fixed $s$ might develop a conjugate point at or after the cut point and hence outside the restricted $\tau$-domain $[0,\tau_-(s))$. We conclude that the image of $\Gamma$ is really a surface.
Observe that since $X$ does not vanish it can even be defined at the cut point. However, by changing the generator ending at the endpoint (and $\Gamma$) one would get a different value of $X$.
This fact will play no significant role in what follows due to the fact that the set of cut points has negligible measure.

So far $X$ has been defined over  the surface defined by the mapping $(s,t)\mapsto\Gamma(s,\tau)$. By taking generators of  $\zJ^-(\gamma(s))$  with starting tangent vector having different components with respect to the base $\{e_\mu\}$ we obtain a vector field defined over $\cup_s \zJ^-(\gamma(s))$.

As the Riemann tensor is Lipschitz, and since $X$ satisfies the Jacobi equation, by using the dependence on initial conditions of first order ODEs we have that the vector field $X$ is Lipschitz, a fact to be used below.

Let $\mcU$ be any relatively compact
domain containing the support of $\varphi$. Consider the map $\hat{L} \colon T\mcM \to \mathbb{R}$ defined as
\[
Z \mapsto \int_{\zJ^-(\pi(Z))\cap \mcU} \varphi  \, L(Z) \rfloor d\mu_g,
\]
where $L(Z)$ is the Jacobi field $X$ over $E^{-}(\pi(Z))$ obtained as the solution of the Jacobi equation  over each generator by imposing, (a)   $L(Z)=Z$ at $\pi(Z)$,  and (b) a vanishing derivative at $\pi(Z)$ in the direction of the affinely parametrised generator: $\frac D {d\tau} L(Z)|_{\tau=0}=0$.

The field $L(Z)$ is linear in $Z$ and hence at each point depends continuously on $Z$.
 Moreover, by
Proposition~\ref{P18XI12.1} the sets $\partial(\zJ^-(p))\cap
\mcU $, when non-empty, are continuous radial graphs which vary
continuously with $p$.  As the domain of integration $\zJ^-(\pi(Z))\cap \mcU$ and the integrand depend continuously on $Z$  we conclude that $\hat{L}(Z)$ is continuous.
 In particular, if $Z(p)$ is a continuous vector field, then $\hat{L}(Z(p))$ is continuous in $p$.

We have the following:

\begin{lemma}
 \label{L18X14.1}
Let $(\mcM,g)$ be globally hyperbolic with a $C^{2,1}$ metric $g$. Suppose that $\varphi$ is smooth and compactly supported.
Then $\tau_\varphi^-$ is differentiable and for every $Z\in T\mcM$ we have
\begin{equation} \label{ert}
Z(\tau^-_\varphi)=\hat{L}(Z).
\end{equation}
\end{lemma}

\begin{proof}
Let us first assume that $Z$ is future-directed timelike and set $p=\pi(Z)$.
Let $\gamma$ be a parametrised
future-directed inextendible timelike curve such that $Z$ is the tangent vector at $p=\gamma(0)$.
Since $\varphi$ is supported in $\mcU$, formula~\eq{17XI12.1} can be rewritten as
\bel{18XI12.11a}
\tp(q) = \int_{J^{-}(q)\cap \mcU} \varphi  \, d\mu_g
 \;,
\ee
for each $q \in \mcM$.
We want to calculate the derivative of $\tp \circ\gamma(s)$ with respect to $s$ at $s=0$, showing in the course of the calculation that this derivative exists.

Denote by $X$ the Jacobi field induced from $\dot{\gamma}$  as explained above.

Suppose, first, that $X$ is Lipschitz in a neighborhood of the support of $\varphi$. Then, at least for small $s$, $ J^-(\gamma(s))\cap \mcU$ is obtained by flowing $J^-(\gamma(0))\cap \mcU$ along $X$. It is then standard that $\tp\circ\gamma$ is differentiable near $s=0$, with
\begin{align}
 Z(\tau^{-}_\varphi)&= \frac{d(\tp\circ\gamma)}{ds}(0) =\int_{J^-(p)\cap \mcU} L_X[\varphi   d\mu_g]
 \label{18XI12.11-2}
 \;.
\end{align}
We can now use the identity $L_X=d i_X+i_X d$ in the integral above. The term $i_X d$ gives a vanishing contribution since $\varphi d \mu_g$ has already maximum degree as a differential form, while the former contribution can be integrated according to
Stokes' theorem for Lipschitz fields on domains with Lipschitz boundaries~\cite[Section~5.8]{EvansGariepy}
to give
\begin{align}
 Z(\tau^{-}_\varphi)&
=\int_{J^-(p)\cap \mcU} L_X[\varphi   d\mu_g] =  \int_{\zJ^-(p)\cap \mcU} \varphi  \, X \rfloor d\mu_g
 \label{18XI12.11-3}
 \;,
\end{align}
as desired.

However, $X$ will not be Lipschitz in general. In fact, in general $X$ will not even extend by continuity to the null cut set. In such cases we proceed as follows:
Let $\Sigma(s)$ be the set at which $\dot J^{-}(\gamma(s))$ fails to be a $C^2$-manifold. It can be useful to recall that $\Sigma(s)=\gamma(s)\cup \Sigma'(s)\cup \Sigma''(s)$, where~\cite{Alberti} (compare~\cite{ChDGH} for a proof of pseudoconvexity of acausal boundaries, as needed to apply~\cite{Alberti}),
in space-time dimension $n+1$, $\Sigma'(s)$ is included in a rectifiable $(n-1)$-manifold, and $\Sigma''(s)$ has vanishing $(n-1)$-dimensional Hausdorff measure.
Let
$$
 \Sigma = \cup_s \Sigma(s)
 \;.
$$
It is well known that $\Sigma\cap \mcU$ has
zero $(n+1)$-dimensional Lebesgue measure, but we give the argument for completeness. For this, let $C^-$ denote the past light cone in Minkowski space-time minus its vertex.
Let
$$
 \Phi \colon \R \times C^- \to \mcM
$$
be the map which to a point $(s, X) \in \R \times C^-$ associates $\exp(X)$,
where $X$ is viewed as a vector in $T_{\gamma(s)} \mcM$ using the construction in the paragraph following \eq{8XI14.5}. Then $\Phi$ is a locally Lipschitz map from $\R \times C^-$ to $\mcM$.

For each $s$ consider the inverse image $\Phi^{-1}(\Sigma(s))$. Now, every null geodesic in $C^-$ intersects the set $\Phi^{-1}(\Sigma(s))$ at at most one point. Fubini's theorem with respect to the measure $\mu_{n}$ induced
on $C^-$ from the Lebesgue measure
using the flat metric on $\R^{n+1}$
shows that $\mu_n(\Phi^{-1}(\Sigma(s)))=0$. This implies that $\cup_s\Phi^{-1}(\Sigma(s))$ is measurable on $\R\times C^-$ with respect to the product measure $\lambda^1\times \mu_n$. Using Fubini's theorem again we obtain
$$
 (\lambda^1\times \mu_n)\big(\cup_s\Phi^{-1}(\Sigma(s))\big) = 0
 \;,
$$
where $\lambda^1$ is the Lebesgue measure on $\R$.
Since $\cup_s \Sigma(s)$ is the image by the locally Lipschitz map $\Phi$ of $\cup_s\Phi^{-1}(\Sigma(s))$,
we conclude that
\bel{8XI14.6}
 \mu_g (\Sigma) = \mu_g(\cup_s\Sigma(s)) =0
\;,
\ee
where $\mu_g$ is the usual metric measure on $\mcM$.

 Using global hyperbolicity it is pretty easy to show that every point $p\in I^{-}(\gamma)$ belongs to one and only one set $E^{-}(\gamma(s))$, $s\in \mathbb{R}$. Thus $I^{-}(\gamma)=\exp (\mathbb{R}N_h\vert_{\gamma})$ where $N_h\vert_{\gamma}$ is the past $h$-unit lightlike bundle over $\gamma$. However, the image of the star domain in which this exponential map is a local diffeomorphism is $I^{-}(\gamma)\setminus\Sigma$, which must be open by local injectivity, thus $\Sigma$ is closed in the topology of $I^{-}(\gamma)$  (the argument is analogous to that used in \cite{kobayashi:nomizu}, Sect. VII.7 vol II, to show that the cut point set is closed). As a consequence for any chosen interval $[\underline{s}, \overline{s}]$, $\bar \Sigma=\cup_{s\in [\underline{s}, \overline{s}]} \Sigma(s)$ is closed.

Let $d$ denote the distance in $\mcM$ from the set $(\mbox{\rm supp}\, \phi) \cap \bar \Sigma$ with respect to our auxiliary complete Riemannian metric $h$. The set $(\mbox{\rm supp}\, \phi) \cap \bar \Sigma$ is closed and compact
which implies that $d$ is Lipschitz. Let $f \colon \R\to \R$ denote any smooth non-decreasing function which vanishes on $ (-\infty, 1/2]$ and equals one on $[1,\infty)$. Set
$$
 \phi_\epsilon = f(d/\epsilon)
\;.
$$
Then $\phi_\epsilon$ is Lipschitz, vanishes in a neighborhood of $\bar \Sigma$, and
$$
 \forall p \not \in \bar\Sigma \quad \phi_\epsilon (p) \to_{\epsilon\to 0} 1
 \;.
$$
Since the vector field $X$, $X=L(Z)$, $Z=\dot\gamma(s)$, $s\in[s_1,s_2]\subset (\underline{s}, \overline{s})$, is Lipschitz on the support of $\phi_\epsilon\varphi$ and this support does not intersect $\bar \Sigma$, we have by the result already established
\bel{8XI14.2}
\tau_{\phi_\epsilon \varphi}^-(\gamma(s_2))-
\tau_{\phi_\epsilon \varphi}^-(\gamma(s_1))  =
 \int_{s_1}^{s_2} \left( \int_{\zJ^-(\gamma(s))} \phi_\epsilon\varphi  \, L(Z) \rfloor d\mu_g \right) ds \;.
\ee
From the dominated convergence theorem we have
$$
 \tau_{\phi_\epsilon \varphi}^-(s)
 \to_{\epsilon\to 0}
 \tau_{\varphi}^-(s)
 \;,
 \quad\int_{\zJ^-(\gamma(s))} \phi_\epsilon\varphi  \, L(Z) \rfloor d\mu_g
 \to_{\epsilon \to 0}
 \int_{\zJ^-(\gamma(s))} \varphi  \, L(Z) \rfloor d\mu_g
 \;.
$$
Passing to the limit $\epsilon\to0$ in \eq{8XI14.2} we obtain
\bel{8XII14.3}
\tau_{\varphi}^-(\gamma(s_2))-
\tau_{\varphi}^-(\gamma(s_1)) =
 \int_{s_1}^{s_2} \left( \int_{\zJ^-(\gamma(s))} \varphi  \, L(Z) \rfloor d\mu_g \right) ds
\;.
\ee
It follows from Lebesgue's continuity theorem that the integrand is a continuous function of $s$. Our derivative formula for timelike $Z$ immediately follows.

It remains to prove the formula for any vector $S\in T\mcM$, namely let us prove %
$$
 \tau^-_\varphi(\exp (S\epsilon))-\tau^-_\varphi(\pi(S))=\epsilon \hat{L}(S)+o(\epsilon)
  \;.
$$
Let   $p=\pi(S)$ and  let $T\in T_p \mcM$ be a future-directed timelike vector such that $T+S$ is future-directed timelike. By continuity we can find a small normal coordinate neighborhood, with coordinates  $\{x^i\}$, such that the vectors $(S^i+T^i)\partial_i$ and $T^i\partial_i$ (constant components) are timelike over the neighborhood. Then
\begin{align*}
    \tau^-_\varphi(S^i\epsilon)-\tau^-_\varphi(0)
        &=\tau^-_\varphi(S^i\epsilon)-\tau^-_\varphi((S^i+T^i)\epsilon))
        +\tau^-_\varphi((S^i+T^i)\epsilon)
    -\tau^-_\varphi(0)
\\
 &=
    -\hat{L}|_{S^i\epsilon}( T^i\partial_i)\epsilon+\hat{L}|_{0}(S+T)\epsilon+o(\epsilon)
\\
 &=
    -\hat{L}|_{0}(T)\epsilon+\hat{L}|_{0}( T+S)\epsilon+[\hat{L}|_{0}(T)-\hat{L}|_{S^i\epsilon} (T^i\partial_i)]\epsilon+o(\epsilon)
\\
 &=\hat{L}|_{0}(S)\epsilon+o(\epsilon),
\end{align*}
where we used the continuity of $p\mapsto \hat{L}|_p$ to infer that the term in square brackets vanishes as $\epsilon\to 0$.
%
\end{proof}

The identity $Z(\tau^{-}_\varphi)=\hat L(Z)$ and the  continuity of $\hat{L}$ imply that $\tau^-_\varphi$ is continuously differentiable.

\begin{remark}
 \label{R11XII12.1}
Both for our purposes here and those of next section, we note that if $\tau$ is a continuously differentiable function such that $C(\tau)>0$ for every future-directed causal vector $C$ then $-\nabla \tau$ is future-directed and timelike. Indeed, $C(\tau)=g(\nabla \tau, C)$ and $g(Y,C)$ is positive for every future-directed causal vector $C$ if and only if $Y$ is past-directed and timelike as can be easily checked in an orthogonal base at the point.
\end{remark}

As such, Remark~\ref{R11XII12.1} implies that $\nabla
\tau^-_\varphi(p)$ is past-directed and timelike provided $E^{-}(p)$
intersects the interior of the support of $\varphi$, namely the open
set $V=\{x:\varphi(x)>0\}$. Indeed, if $\gamma$ is a causal curve
the integrand in $\hat{L}(\dot \gamma)$  reads
\[
 \varphi  \, X \rfloor d\mu_g
 =\varphi \,g(X,\frac{d}{d \tau})\,  d A\, d \tau
  \; ,
\]
where $X$ in the Jacobi field induced from $\dot \gamma$, $dA$ is the area element transverse to the generators of
$E^{-}(\gamma(s))$. In order to show that the integral is positive
recall, from above, that  $g(X,\frac{d}{d \tau})$ is constant over
$\Gamma_s(\tau)$. Hence it coincides with its value at the tip
$\gamma(s)$, where it is  $g(\dot{\gamma},\frac{d}{d \tau})$. As
$\gamma$ is causal,
 and $\frac{d}{d \tau}$ is null and
past-directed, this scalar product  is positive unless
$\dot{\gamma}(s)$ is null and $\frac{d}{d \tau}$ is proportional to
it. However, as the integral involves all directions, for $n\ge 2$ this
exceptional null generator does not affect the positivity of the
integral, as it has vanishing measure within $\zJ^-(\gamma(s))\cap
\mcU$. The conclusion does not change for $n=1$ since the integral would be the sum of the non-negative contribution from two lightlike geodesic segments and only one of those can vanish.

Stated in another way, if $E^{-}(\gamma(s))$ intersects $V$, since
$V$ is open we can always find a generator of $E^{-}(\gamma(s))$ not
aligned with $X$ at $\gamma(s)$ and intersecting $V$. The integral
in a neighborhood of this generator gives a positive contribution.
Thus either $E^{-}(\gamma(s))$ does not intersect $V$ and $\nabla
\tau^-_\varphi(p)$ vanishes, or $E^{-}(\gamma(s))$ intersects $V$
and $\nabla \tau^-_\varphi(p)$ is timelike and past directed.

\medskip

Thus, we have proved:

\begin{lemma}
 \label{L18XI12.1}
In globally hyperbolic spacetimes the functions $\tpm$ are
continuously differentiable with timelike or vanishing gradient for
all continuous compactly supported non-negative functions $\varphi$.
 \qed
\end{lemma}

However, $\tp$ is zero on $\mcM\setminus J^{+}(\mathrm{supp}(\varphi))$, so it is not a time function there. Similarly, $\tp$ is constant near every point $p$ such that $\mathrm{supp} (\varphi) \subset I^-(p)$. So, a little more work is needed to construct a differentiable time function:

Let $\{B_{p_i}(r_i)\}_{i\in\N}$ be any locally finite
covering of $\mcM$ with open $h$-balls centred at $p_i$ with
$h$-radius $r_i\le 1$. Let $\varphi_i$ be a partition of unity
associated with this covering. Let $\tpi$ be the associated
(continuously differentiable) volume functions. Define
$$
 c_i=\sup_{q\in\mcM} \tpi (q)
 \;,
 \qquad
 C_i = 1+ c_i +\sup_{p\in B(p_1,i)} |D \tpi|_h < \infty
 \;.
$$
\newcommand{\myBi}{D_i}%
For any sequence $\myBi\ge C_i$ set
\begin{equation} \label{akc}
 \varphi = \sum_i \frac{1}{2^i\myBi}\, \varphi_i
\end{equation}
(in what follows the reader can simply assume that $\myBi=C_i$, the point of introducing the $\myBi$'s is to make it clear that any  sequence $\{D_i\}$ with $\myBi\ge C_i$  leads to a differentiable time function).
Consider the function
\bel{18XI12.13}
 \tp(p) = \int_{{J}^{-}(p) } \varphi  \, d\mu_g = \sum_i \frac{1}{2^i\myBi}\, \tpi
 \;.
\ee
Let $K$ be a compact subset of $\mcM$, there exists $n\in \N$ such that $K\subset B_{p_1}(n)$.
Then
\bean
\sup_{q\in B_{p_1}(n)} \sum_{i=1}^\infty \frac{1}{2D_i} \vert
D\tau_{\varphi_i}^-\vert_h & \le & \underbrace{ \sum_{i=1}^n
\frac{1}{2^i\myBi} \sup _{q\in B_{p_1}(n)}
    |D\tpi|_h(q)}_{<\infty}
 + \sum_{i=n+1}^\infty
 \frac{1}{2^i\myBi}\,  \underbrace{\sup _{q\in B_{p_1}(n)} |D\tpi|_h(q)}_{\le C_i}
 \nonumber
\\
 & < & \infty
 \;.
\label{18XI12.21}\end{eqnarray}
This shows that the series defining $\tp$ converges in $C^1$ norm on
every compact set, resulting in a differentiable function. Since
each $\tpi$ has timelike or vanishing gradient, with $d\tpi$
non-vanishing on the interior of $\mathrm{supp}(\tpi)$, the
timelikeness of $\nabla \tp$ readily follows, and
Theorem~\ref{T15XI12.1} is proved.
\qed

\section{Smoothing anti-Lipschitz time functions}
 \label{s18XI12.1}

In this section we first show that the volume time functions of the
previous section are locally anti-Lipschitz, a property to be
defined shortly, and then that any time function which shares the
anti-Lipschitz property with respect to a  metric with wider light
cones can be smoothed. These results are then applied to prove the
existence of smooth time functions in stably causal spacetimes, and
smooth Cauchy time functions in globally hyperbolic spacetimes.
Finally, using the equivalence between stable causality and
$K$-causality we prove that the existence of a  time function implies
the existence of a smooth one.

We begin with a simple lemma.

\begin{lemma}
Let $(\mcM,g)$ be a strongly causal spacetime.   The following two conditions on a  function $\tau^+ \colon \mcM \to \mathbb{R}$,
 respectively $\tau^- \colon \mcM \to \mathbb{R}$, are equivalent:
\begin{itemize}
\item[(i)] for every point $p\in \mcM$ there exists a relatively compact neighborhood $\mcO_p$ of $p$ and a constant $C_p>0$ so that for every $h$-parametrised past-directed (resp.\ future-directed) causal curve $\gamma$ with image in $\mcO_p$ we have, for all $s_2\ge s_1$,
\bel{12XII12.1}
 \nohattau^\pm(\gamma(s_2))- \nohattau^\pm(\gamma(s_1))\ge C_p (s_2-s_1)
 \;,
\ee
\item[(ii)] for every compact set $K$ there is a constant $C_K>0$ such that for every $h$-parametrised past-directed (resp.\ future-directed) causal curve $\gamma$ with image in $K$, $\tau^+$ (resp.\ $\tau^-$) satisfies, for all $s_2\ge s_1$,
\bel{12XII12.1b}
 \nohattau^\pm(\gamma(s_2))- \nohattau^\pm(\gamma(s_1))\ge C_K (s_2-s_1)
 \;.
\ee
\end{itemize}
\end{lemma}
Clearly, both conditions imply that $\mp\tau^\pm$ is a time function.

{\noindent \sc Proof:}
$(ii) \Longrightarrow (i)$. Just take  the relatively compact neighborhood to be the interior $\mathring K$ of any compact set $K$ so that $p$ belongs to $\mathring K$.

$(i) \Longrightarrow (ii)$.  Since $(\mcM,g)$ is strongly causal, each point $p\in K$ belongs to a relatively compact open causally convex set $\hat\mcO_p\subset \mcO_p$, thus there is a finite subcovering of $K$, $\{\hat\mcO_{p_j}: j=1,\cdots,n\}$. Since no causal curve can enter  a causally convex set twice, (ii) holds with $C_K:=\min_j C_{p_j}$.
\qedskip

We shall say that $\tau$ is  {\em locally ($\pm$-$g$-)anti-Lipschitz} if it satisfies (i) or (ii) above. Clearly, this property is independent of the Riemannian metric $h$ used, as two different Riemannian metrics are Lipschitz equivalent over compact sets. (In space-times which are not strongly causal, one could use e.g. \eq{12XII12.1} as a definition of anti-Lipschitz in general, but this generality will not be needed in what follows.)

\begin{remark}
 \label{R15X14}
Observe that a past volume function can be discontinuous and yet locally anti-Lipschitz, e.g.\ remove a past inextendible timelike geodesic, including the future endpoint, from a strip $(-1,1)\times \mathbb{R}$ of Minkowski 1+1 spacetime with coordinates $(t,x)$.
\end{remark}

\begin{proposition} \label{bya}
Let $\tau^- \colon \mcM \to \mathbb{R}$ be continuously differentiable. Then $\tau^-$ has past-directed timelike gradient if and only if it is locally anti-Lipschitz. \end{proposition}

There is evidently a time-dual version of Proposition~\ref{bya}.

\medskip

{\noindent \sc Proof:}
Suppose that $\tau^-$ is anti-Lipschitz. Let $X$ be a $h$-normalized future-directed causal vector at $p$, and let $\gamma(s)$ be a causal curve with tangent $X$ at $p=\gamma(0)$. Taking the limit $s\to 0$ of the anti-Lipschitz condition we find $X(\tau^-)\ge C_K>0$ where $K$ is a compact neighborhood of $p$. Since $X$ is arbitrary, using Remark \ref{R11XII12.1} we infer that $\nabla \tau$ is past-directed and timelike.

Conversely, let us assume that  $\tau^-$ has past-directed timelike
gradient,  and let $K$ be a compact set. Let us observe that $ d
\tau=g(\nabla \tau, \cdot)$. Let $g_\parallel=\frac{1}{g(\nabla
\tau,\nabla \tau)} \, d \tau^2$, and let $g_\perp$ be a quadratic
form such that $g=g_\perp+g_\parallel$. Let $\tilde g=\alpha g_\perp
+g_\parallel=\alpha g+(1-\alpha) g_\parallel$, with $\alpha <1$. For
$\alpha$ sufficiently close to 1, $\tilde g$ is Lorentzian over $K$
with light cones wider than those of $g$, and moreover $\tilde
g(\nabla \tau, X)=g(\nabla \tau, X)$ for $X\in TK$. Let
 $X\in TK$ be any $h$-normalized future-directed causal vector, then
$ \tilde g(X,X)<0$ and there is $C_K>0$ such that
$$
 X(\tau^-)=g(\nabla \tau^-,X)=\tilde g(\nabla \tau^-,X) \ge [ \tilde g(\nabla \tau^-,\nabla \tau^-) \tilde g(X,X)]^{1/2}\ge C_K
 \;,
$$
where in the last inequality we used the compactness of the bundle of $h$-normalized causal vectors in $TK$. From here the anti-Lipschitz condition follows upon integration in $h$-arc length $s$.
\qed

\begin{corollary}
Let $(\mcM, g)$ be globally hyperbolic.   The  continuously differentiable function $\tau_\varphi^-$ of  Theorem \ref{T15XI12.1}, with   $\varphi >0$,
 is locally anti-Lipschitz.
\end{corollary}

Actually we can prove something more. We shall  need a simple preliminary result:

\begin{lemma} \label{nox}  Let $f \colon \mathbb{R}\to\mathbb{R}$ be a non-decreasing function such that for
every $s\in \mathbb{R}$,
$$\liminf_{\epsilon\to 0^+}
\frac{1}{\epsilon} [f(s+\epsilon)-f(s)]\ge C\ge  0 \; ,$$ then if
$s_2-s_1\ge 0$, we have $f(s_2)-f(s_1)\ge C(s_2-s_1)$.
\end{lemma}

{\noindent \sc Proof:}
The assumption for $s=s_1$ tells us that there is a maximal
right-neighborhood $[s_1,\tilde{s})$ such that for every $s\in
[s_1,\tilde{s})$, we have $f(s)-f(s_1)\ge C(s-s_1)$. Let us show
that $\tilde s$ must be infinite and thus that $s_2\in
[s_1,\tilde{s})$. For, if not, taking the limit for $s\to \tilde{s}$
of $f(s)-f(s_1)\ge C(s-s_1)$, and using the fact that $f$ is
non-decreasing, we obtain $f(\tilde s)-f(s_1)\ge C(\tilde s-s_1)$.
But from the assumption applied to $\tilde s$, there is
$\hat{s}>\tilde s$ such that for $s \in [\tilde s, \hat s)$,
$f(s)-f(\tilde s)\ge C(s -\tilde s )$, which summed to the previous
equation gives $f( s)-f(s_1)\ge C( s-s_1)$, for $s\in [s_1, \hat
s)$, showing that $\tilde s$ was not maximal, a contradiction.
\qed

\begin{Theorem} \label{nju}
Let $(\mcM,g)$ be past-distinguishing   where $g$ is $C^{2,1}$,   and  let
$\tau_\varphi^{-}$ of \eq{17XI12.1} be  defined through a
continuous function $0<\varphi\in L^1(\mcM)$. Then
$\tau_\varphi^{-}$  is locally anti-Lipschitz.
\end{Theorem}

We emphasise that $\tau_\varphi^{-}$ might not   be continuous without further hypotheses, compare Remark~\ref{R15X14}.

\medskip

{\noindent \sc Proof:}
We just need to show that for every $p\in \mcM$ there is a
neighborhood $U$, and a positive continuous function $c: U \to
(0,+\infty)$ such that if $\gamma$ is a future-directed $h$-arc
length parametrised causal curve in $U$, then for every
$q=\gamma(s)$ we have
\[
\liminf_{\epsilon\to 0^+} \frac{1}{\epsilon}
[\tau_\varphi^-(\gamma(s+\epsilon))-\tau_\varphi^-(\gamma(s))]\ge c(q)>  0\; .
\]
By Lemma \ref{nox}, $\tau^-_\varphi$ would be anti-Lipschitz on that
open subset of  $U$ for which $c(q)>c(p)/2$ (with anti-Lipschitz
constant $C=c(p)/2$), and hence, given the arbitrariness of $p$,
$\tau_\varphi^-$ would be locally anti-Lipschitz.

Now, observe that we can find $r>0$, sufficiently small, so that the
ball $B_h(p,r)$ is contained in a past-distinguishing neighborhood
contained in a convex neighborhood contained in a globally hyperbolic neighborhood, so that for every $q\in
B_h(p,r)$, the intersection of $\dot J^-(q)$ with the ball of radius
$r$, $B_h(p,r)$, is a smooth null hypersurface except at the tip
$q$. Let $\phi_p$ be a smooth non-negative ``cut-off" function such
that $\phi_p\le 1$, $\phi_p(q)=1$ for $q\in B_h(p,r/2)$
and with
support in $U:=B_h(p,r)$. Let
\[
c(q):=\inf _{X(q)} \int_{B_h(p,r )\cap \dot J^-(q) }  \phi_p\,
\varphi X \rfloor \, d\mu_g,
\]
where $X$ is the already introduced Jacobi field which depends only
on the $h$-normalized future-directed causal vector $X(q)$ in a
continuous way. As already explained, the integrand is non-negative
when the formula is rewritten in terms of the coordinate-Lebesgue
measure, and the integral is positive and continuous. By
construction, $c(q)$ is then continuous and positive.

Finally, observe that for  $\gamma$ and $q$ as above, if we set
\[
f_q(s):= \int_{B_h(p,r)\cap J^-(\gamma(s))} \phi_p  \,\varphi \,
d\mu_g,
\]
then since $\phi_p$ is supported in a globally hyperbolic neighborhood we can apply the formula of the previous section    $\frac{d }{d s}f_q(s)= \int_{B_h(p,r )\cap \dot J^-(q) }
\phi_p \, \varphi  X \rfloor \, d\mu_g \ge c(q)$ and
\[
\tau_\varphi^-(s+\epsilon)-\tau_\varphi^-(s)=\int_{J^-(\gamma(s))\backslash
J^{-}(q)}  \varphi \, d\mu_g \ge \int_{B_h(p,r)\cap
J^-(\gamma(s))\backslash J^{-}(q)} \phi_p \, \varphi \, d\mu_g=
f_q(s+\epsilon)-f_q(s),
\]
from which we obtain the desired conclusion.
\qedskip

\begin{remark}
In the   proof above we used the derivative formula for the volume function which we obtained in the previous section. In this application we are working in a convex neighborhood contained in a globally hyperbolic neighborhood since the argument of the integral includes a cut-off function $\phi_p$. In the current setting the proof of the derivative formula is in fact much simpler as there are no  focusing or cut points to $p$ in the supports of $\phi_p$.
\end{remark}

We recall that $\hat{g}\succ g$ means that the causal cone of $g$ is
contained in the timelike cone of $\hat{g}$ at all points in
spacetime. If we can find $\hat g$ such that $(\mcM,\hat g)$ is
causal, then we say that $(\mcM,g)$ is stably causal. We also recall
that a \emph{Cauchy time function} is a time function onto
$\mathbb{R}$ whose level sets are intersected (precisely) once by
every inextendible causal curve. A spacetime admits a Cauchy time
function if and only if it is globally hyperbolic
\cite{GerochDoD,HE}.

\begin{Theorem}
 \label{T16IV12.1} Let $(\mcM,g)$ be a stably causal spacetime with a continuous metric $g$, and
 let $\nohattau$ be a time function on $\mcM$. Moreover, suppose that
  \begin{itemize}
      \item[(*)]  there exists a metric $\hg\succ g$  such that $\tau$ is locally $\hat{g}$-anti-Lipschitz.
  \end{itemize}
Then for every function $\alpha: \mcM \to (0,+\infty)$ there exists
a smooth $g$-time function $\hat{\tau}$, with $g$-timelike gradient,
such that $\vert \hat\tau-\nohattau\vert <\alpha$.
 As a consequence, if $\tau$ is Cauchy we can choose  $\hat \tau$  Cauchy (take $\alpha$ bounded).
\end{Theorem}

{\noindent \sc Proof:}
Consider $p\in \mcM$, let $x^\mu$ be local coordinates near $p$, and
let $\mcC_{g,p}\subset T_pM$ denote the collection of $g$-causal
vectors at $p$. By continuity, there exists $\epsilon(p)> 0$ so that
for all $q$, $q'$ in a relatively compact coordinate ball
$B_p(3\epsilon(p))$ of radius $3\epsilon(p)$ centred at $p$ and for
all vectors $X(q)=X^\mu(q)\partial_\mu|_q \in \mcC_{g,q}$ the vector
$X(q'):=X^\mu(q)\partial_\mu|_{q'} \in T_{q'}\mcM$, with
coordinate components $X^\mu(q')$ at $q'$ equal to its coordinate components $X^\mu(q)$ at $q$,
is $\hg$-timelike at $q'$. The constant $\epsilon$ can be chosen so small
that if $X\in T_q \mcM$ and $Y\in T_{q'} \mcM$ are two non-zero vectors on
$T {B_p(3\epsilon(p))}$ such that $X^\mu(q)=Y^\mu(q')$, then the
 ratio of their $h$-norms belongs to $[1/2,2]$.

Let $\{\mcO_i:={B_{p_i}(\epsilon_i)}\}_{i\in\N}$ be a locally finite
covering of $\mcM$ by such balls.
Let $\varphi_i$ be a partition of unity subordinate
to the cover $\{\mcO_i\}_{i\in\N}$. Choose some
$0<\eta_j<\epsilon_j$. In local coordinates on $\mcO_j$ let $\tau_j$
be defined by convolution with an even
non-negative function $\chi$,
supported in the coordinate ball of radius one, with integral one:
$$
 \tau_j (x) =\left\{
               \begin{array}{ll}
 \frac 1 {\eta_j^{n+1}} \int_{B_{p_j}(3\epsilon_j)} \chi\left(\frac{y-x}{\eta_j}\right) \nohattau (y)\, d^{n+1}y, & \hbox{$x\in B_{p_j}(2\epsilon_j)$;} \\
                 0, & \hbox{otherwise.}
               \end{array}
             \right.
$$
We define the smooth function
$$
 \htau:= \sum_j \varphi_j \tau_j
 \;.
$$
The non-vanishing terms at each point are finite in number, and
$\hat\tau$ converges pointwise to $\tau$ as we let the constants
$\eta_j$ converge to zero. The idea is to control the constants
$\eta_j$ to get the desired properties for $\hat\tau$.

Let $x\in \mcO_j=B_{p_j}(\epsilon_j)$, and let $X = X^\mu
\partial_\mu$ be any $g$-causal vector at $x\in \mcO_j$, of
$h$-length one, then the curve $x^\mu (s)= x^\mu + X^\mu s$ is
$\hg$-timelike
 as long as it stays within $B_{p_j}(3\epsilon_j)$. We observe that $s$ is not the $h$-arc length parametrization of the curve, however from our choice of $\epsilon$ we have $\inf_{\overline{\mcO_j}} \vert X^\mu\partial_\mu \vert_h\ge 1/2$,
thus for $s_2>s_1$ it holds that $(s_2-s_1)<2(t_2-t_1)$, where $t$
is the $h$-arc length parametrization. Let $2C_j$ be the
$\hat{g}$-anti-Lipschitz constant over $\overline{\mcO_j}$.

We write:
\beaa
 \htau(x(s))-\htau(x)= \underbrace{\sum_j \big(\varphi_j (x(s))-\varphi_j (x)\big)\tau_j(x(s))}_{=:I(s)} +
  \underbrace{
  \sum_j \varphi_j (x )\big(\tau_j(x(s))-\tau_j(x)\big)
   }_{=:II(s)}
   \;.
\eeaa
We have at $x\in \mcO_j$,
\beaa
 \lim_{t\to 0} \frac{II (s(t)) }{t}&=& \frac{1}{\vert X(x)\vert_h}\lim_{s\to 0} \frac{II (s) }{s}
     =
  \lim_{s\to 0} \frac 1 s
  \sum_k \varphi_k (x )\big(\tau_k(x+ X s)-\tau_k(x))
\\
& = &
  \lim_{s\to 0} \frac 1 s
  \sum_k \frac {\varphi_k (x )}{\eta_k ^{n+1}}\int_{B_{0}(\epsilon_k)} \chi\left(\frac{z}{\eta_k}\right) \big( \underbrace{\nohattau (x+Xs +z)-\nohattau (x+z)}_{\ge C_k s}\big)
   \, d^{n+1}z
\\
& \ge  &
  \sum_k  {\varphi_k (x )} C_k \ge \min_{k\,:\,\mcO_k\cap \mcO_j\ne \emptyset} C_k=:B_j >0
  \;,
\eeaa
where the constant $B_j$ does not depend on the set of constants
$\{\eta_i\}$.

For every $j$ let
\[
R_j:=\sup_{k\,:\,\mcO_k\cap \mcO_j\ne \emptyset}\sup_{x\in
\overline{\mcO_j}} \vert \nabla^h \varphi_k(x)\vert_h\; ,
\]
let $N_j$ be the number of distinct sets $\mcO_k$ which have
non-empty intersection with $\mcO_j$, and let us choose  $\eta_j$ so
small that
\[
\sup_{x\in \overline{\mcO_j}} |\tau(x)-\tau_j(x)|<
\min_{\ell:\mcO_\ell\cap \mcO_j\ne \emptyset} \{\,\frac{1}{N_\ell}
\inf_{\overline{\mcO_\ell}}\alpha, \,\frac{B_\ell}{2N_\ell
R_\ell}\}\; .
\]
Let $\chi_k$ be the characteristic function of $\mcO_k$, so that
$\varphi_k\le \chi_k$.  The sets $\mcO_j$ and
$\overline{\mcO_j}$ intersect the same sets of the covering
$\{\mcO_i\}$, which are $N_j$ in number, thus
\[
 \sup_{x\in \overline{\mcO_j}}  \sum_{k:\mcO_k\cap
\mcO_j\ne \emptyset}[\chi_k(x) |\tau(x)-\tau_k(x)|]\le \! \!\!
\sum_{k:\mcO_k\cap \mcO_j\ne \emptyset}\, \sup_{x\in
\overline{\mcO_k}}|\tau(x)-\tau_k(x)|\le\!\!\!
\sum_{k\,:\,\mcO_k\cap \mcO_j\ne \emptyset} \frac{B_j}{2R_j N_j}=
\frac{B_j}{2R_j}
 \;.
\]


Then at $x\in \mcO_j$,
\beaa \bigg | \lim_{t\to 0} \frac{I (s(t)) }{t}\bigg |&=&\bigg |
\lim_{s\to 0} \frac{I (s) }{s} \bigg|
     = \bigg |
  \lim_{s\to 0} \sum_k \frac{\varphi_k (x(s))-\varphi_k (x)} s\, \tau_k(x(s))
 \bigg|
\\
& = &\bigg |
  \sum_k X\big(\varphi_k(x)\big) \tau_k (x )\bigg |
= \bigg |\sum_k X\big(\varphi_k(x)\big) \big[\tau(x)-\big(\tau(x)
-\tau_k (x )\big)\big]
 \bigg|
\\
 &= & \bigg |\underbrace{X\bigg(\sum_k \varphi_k(x) \bigg)}_{=  X(1)  = 0}\tau (x )
- \sum_k X\big(\varphi_k(x)\big) (\tau(x) -\tau_k (x ) )
 \bigg|
\\
 & \le & \sum_k |X\big(\varphi_k(x)\big)| \,|\tau(x) -\tau_k (x )|=\sum_{k\,:\,\mcO_k\cap \mcO_j\ne \emptyset} |X\big(\varphi_k(x)\big)|\, |\tau(x) -\tau_k (x )|
\\
 &\le & R_j \sum_{k:\mcO_k\cap \mcO_j\ne \emptyset}  \chi_k(x) |\tau(x) -\tau_k (x )|\le \frac{B_j}{2}
  \;.
\eeaa
Hence, for   every $x\in \mcO_j$ and every $g$-causal vector $X\in
T_x\mcM$ of $h$-length one, there exists a constant $B_j/2$ such
that we have
\bel{1IV12.1}
 X(\hat\tau) \ge B_j/2
 \;.
\ee

In particular $\hat\tau$ is a differentiable function which is
strictly increasing along any $g$-causal curve. By Remark~
\ref{R11XII12.1}, the $g$-gradient of $\hat\tau$ is everywhere
$g$-timelike. Finally, for every $x\in \mcM$, there is some $j$ such
that $x\in \mcO_j$, hence
\begin{align*}
\vert \tau(x)-\hat\tau(x)\vert&=\vert \sum_k \varphi_k(x)[\tau(x)-\tau_k(x)] \vert\le \sum_{k:\mcO_k\cap \mcO_j\ne \emptyset} \sup_{x\in \overline{\mcO_k}} |\tau(x)-\tau_k(x)|\\
&\le \sum_{k:\mcO_k\cap \mcO_j\ne \emptyset} \frac{1}{N_j}
\inf_{\overline{\mcO_j}} \alpha\le \alpha(x) \sum_{k:\mcO_k\cap
\mcO_j\ne \emptyset} \frac{1}{N_j}= \alpha(x) \; .
\end{align*}

Note that the smoothness of $\hat\tau$ depends only upon the
smoothness of $\mcM$, regardless of the smoothness of the metric.

For the last claim, since $\tau$ is Cauchy it is onto
$\mathbb{R}$ thus the same holds for $\hat \tau$, and since each constant
slice $\hat \tau^{-1}(t)$ is contained in $\tau^{-1}([t-1,t+1])$,
namely between the Cauchy hypersurfaces $\tau^{-1}(t-1)$ and
$\tau^{-1}(t+1)$, the level-set $\hat \tau^{-1}(t)$ is also a Cauchy
hypersurface, and thus $\hat \tau$ is Cauchy.
\qed

\medskip

In a distinguishing spacetime  the functions $\tau^-_\varphi$ and
$\tau^+_\varphi$, though increasing over future-directed (resp.\
past-directed) causal curves, might be only upper semi-continuous and thus might
fail to be time functions. Indeed, they are continuous if and
only if the spacetime is causally continuous
\cite{HawkingSachs,MinguzziSanchez}.
 Under the weaker notion of stable causality
Hawking was able to construct a time function averaging Geroch's
volume functions
for wider metrics~\cite{HawkingPRSL68,HE}, as
follows: Suppose that $(M,g)$ is stably causal, so that there is
$\tilde g\succ g$ such that $(M,\tilde g)$ is causal. Without loss
of generality we can assume $\tilde g$ to be $C^2$. Let
\[
g_\lambda=(1-\frac{\lambda}{3})g+\frac{\lambda}{3} \,\tilde g,
\qquad \lambda \in [0,3].
\]
Clearly, $g_0=g$, $g_3=\hat{g}$ and if $\lambda_1<\lambda_2$ then
$g_{\lambda_1} \prec g_{\lambda_2}$. In particular, for each
$\lambda$, $(M,g_\lambda)$ is causal.

Let $\mu$ be a finite measure, e.g.\ $d  \mu= \varphi \, d
\mu_{g_0}$, and let us define the Geroch's volume functions
\[
\tau^-_\lambda(p)= \int_{J_{g_\lambda}^{-}(p)} \,  d \mu \; .
\]
Hawking considers the average
\[
\tau^-_H(p)=\int_1^2 \tau^-_\lambda(p) \, d \lambda,
\]
and proves that this function is indeed a time function.

The next result with its corollary  provides the simplest proof that stably causal
spacetimes admit smooth time functions, and that, in fact, they can
be chosen to approximate Hawking's time (previous existence proofs did not establish this approximation property  \cite{BernalSanchez}).
 This result was announced
long ago by Seifert~\cite{Seifert} (with a not-entirely-transparent proof)
 and has been used by Hawking and Ellis
\cite[Prop.\ 6.4.9]{HE} (who referred to Seifert's original doctoral thesis). Our approach is
quite close in spirit to Seifert's original work. We emphasise that
Seifert's article contains many important ideas. In particular,
Seifert was the first to recognize the role of the local
anti-Lipschitz condition (Seifert speaks of {\em uniform} time
functions).

\begin{Theorem} \label{kku}
Let $(\mcM,g)$ be a stably causal spacetime with a $C^{2,1}$ metric $g$.
For every function $\alpha \colon \mcM \to (0,+\infty)$ there exists a
smooth time function ${\tau}^-_\alpha$, with timelike gradient, such
that $\vert \tau^-_H- \tau^-_\alpha\vert <\alpha$.
\end{Theorem}

{\noindent \sc Proof:}
According to Theorem~\ref{T16IV12.1} we need only to prove that
$\tau^-_H$ is locally anti-Lipschitz with respect to $g_{1/2}$. As
we chose $\hat g$ to be $C^{2,1}$ (this can always be  done) we have
that $g_\lambda$ is $C^{2,1}$ with respect to $x\in \mcM$ and $C^\infty$
with respect to $\lambda$. We wish to prove that for $p\in \mcM$ the
functions $\tau^-_\lambda(q)$ are anti-Lipschitz  over a
neighborhood $V\ni p$, with anti-Lipschitz constants $C_\lambda$
that can be chosen to
depend continuously on $\lambda$. If so, since every $h$-parametrized $g_{1/2}$-causal  curve is a $g_\lambda$-causal
curve for $\lambda \in [1,2]$,
 $\tau^-_H$ would be anti-Lipschitz
over $V$ with anti-Lipschitz constant not smaller than $C:=\int_1^2 C_\lambda \,
d\lambda>0$. Indeed, for $s_2\ge s_1$,
\[
\tau^-_H(\gamma(s_2))-\tau^-_H(\gamma(s_1))=\int_1^2[\tau^-_\lambda(\gamma(s_2))-\tau^-_\lambda(\gamma(s_1))]\,
d\lambda\ge  \int_1^2C_\lambda(s_2-s_1)\, d\lambda= C (s_2-s_1).
\]
The fact that $C_\lambda$ is continuous in $\lambda$ follows immediately from continuity in $\lambda$ of  the function $c_\lambda(q)$ mentioned in
Theorem~\ref{nju}: Indeed,  this function reads
\[
c_\lambda(q):=\inf _{X^{(\lambda)}(q)} \int_{B_h(p,r )\cap \dot
J_{g_\lambda}^-(q) } \phi_p\, \varphi X^{(\lambda)} \rfloor \,
d\mu_g,
\]
where $X^{(\lambda)}$ is the Jacobi field obtained by solving the
$g_\lambda$-Jacobi equation. The results on the dependence with
respect to the initial conditions and  parameters of the
theory of ordinary differential equations assure that this function
is continuous~\cite{hartman2}.
\qedskip

\begin{corollary} \label{cos}
Every stably causal spacetime endowed with a continuous metric $g$ admits a smooth time function with
timelike gradient.
\end{corollary}

{\noindent \sc Proof:}
Any stably causal $C^0$ metric $g$ admits some smooth $\hat g\succ g$ such that $(\mcM, \hat g)$ is stably causal. This result holds for $g$ continuous, see \cite{FathiSiconolfiTime} (alternatively, the result might  be obtained following \cite{NavarroMinguzzi} and adapting some steps to the low differentiability case wherever required).
But any smooth time function for $(\mcM, \hat g)$ is a smooth time function for $(\mcM, g)$, which by Remark \ref{R11XII12.1} has timelike gradient with respect to both metrics.
\qedskip

We can use a strategy quite similar to that followed above to prove
existence of smooth Cauchy time functions in globally hyperbolic
spacetimes. This result was also announced  by Seifert~\cite{Seifert,HE} who provided a non-transparent argument. A first detailed proof appeared in~\cite{BernalSanchez1, BernalSanchez}.

\begin{Theorem}
Every globally hyperbolic spacetime $(\mcM,g)$ where $g$ is continuous   admits a smooth Cauchy time function with timelike gradient.
\end{Theorem}

{\noindent \sc Proof:}
Recall that  global
hyperbolicity is stable~\cite{FathiSiconolfiTime,NavarroMinguzzi},
in the sense that it is possible to find a smooth metric $\hat
g\succ g$ such that $(\mcM,\hat g)$ is globally hyperbolic.
So let $\hat g\succ g$ be such that
$(\mcM,\hat g)$ is globally hyperbolic and $\hat{g}$ is smooth. Geroch's time functions
$\tau^+_\varphi$ and  $\tau^-_\varphi$ for the spacetime $(\mcM,\hat g)$ are locally anti-Lipschitz
with respect to $\hat g$. As a consequence $\tau=\ln
(\tau^-_\varphi/ \tau^+_\varphi)$
is also locally anti-Lipschitz with respect to $\hat g$ for some
choice of $\varphi$ (for $\tau^+_\varphi$ can be chosen continuously
differentiable and hence locally  Lipschitz by Theorem \ref{T15XI12.1}). Since $\tau$ is Cauchy for $(\mcM,\hat g)$ it is also Cauchy for $(\mcM,g)$.
The claim follows from the last statement of Theorem~\ref{T16IV12.1}.
\qedskip

An alternative proof, that does not invoke the stability of global
hyperbolicity, will be given in the next section.

We end this section by proving that existence of a time function
implies   existence of a smooth one with timelike gradient. This result was first proved  in \cite{BernalSanchez,sanchez05b} by different methods.
Let us
recall  that $K^+$ is the smallest closed and transitive relation
which contains the causal relation $J^+$. A spacetime is said to be
$K$-causal if $K^+$ is a partial order~\cite{SorkinWoolgar}. A self
contained proof of the equivalence between $K$-causality and stable
causality can be found in~\cite{MinguzziKCausality}.

\begin{Theorem} \label{pxr}
 Let $(M,g)$ be any spacetime with a $C^2$ metric $g$. The following
conditions are equivalent:
\begin{itemize}
\item[(a)] $(M,g)$ admits a
 time function,
\item[(b)] $(M,g)$ admits a smooth time function with timelike gradient,
\item[(c)] $(M,g)$ is stably causal.
\end{itemize}
\end{Theorem}

{\noindent \sc Proof:}
The implication (c) $\Longrightarrow$ (b) is given by Corollary
\ref{cos}. The implication  (b) $\Longrightarrow$ (a) is obvious.
Finally, in order to prove that (a) $\Longrightarrow$ (c), we recall
that any spacetime which admits a time function is $K$-causal
\cite[Lemma 4-(b)]{MinguzziTime}, and $K$-causality coincides with
stable causality~\cite{MinguzziKCausality}.
\qedskip

The previous result probably holds already for continuous metrics $g$ since the proofs given in \cite{MinguzziKCausality,MinguzziTime} do not seem to depend in any essential way on the differentiability of the metric, but we have not attempted to check all details of this.

As shown in~\cite{MinguzziTime} one could go in the other direction,
namely use any independently obtained proof of the implication (a) $\Longrightarrow$
(b) to show the equivalence between $K$-causality and stable
causality.

\section{Extending the anti-Lipschitz property  to wider metrics}
 \label{ss16XII12.1}

The anti-Lipschitz condition with respect to a wider metric $\hat
g\succ g$ is the key ingredient to our  Theorem~\ref{T16IV12.1} on
uniform approximation. Assuming a local Lipschitz condition, we can infer this
property from the anti-Lipschitz condition with respect to $g$. This
allows us to smooth directly the differentiable time functions
$\tpm$ of $(\mcM,g)$ obtained in Section~\ref{hgc}.

We shall need the following lemma:

\begin{lemma}
 \label{L10X14.1}
Let $(Q,h)$ be a Riemannian space, $f$ a locally Lipschitz function,
and let $\gamma \colon [0,1] \to Q$ be an injective $C^2$ curve. We can
find another $C^2$ curve $\alpha \colon [0,1] \to Q$, arbitrarily close to
$\gamma$ in $C^{2}$ norm, such that the differential $d f$ of $f$ exists almost everywhere
on the image of $\alpha$, $f\circ \alpha$ is almost everywhere
differentiable, $\frac{d}{d t}(f\circ \alpha)=d f[\dot\alpha]$
almost everywhere on $[0,1]$ and $f(\alpha(1))-f(\alpha(0))=\int_0^1
d f[\dot\alpha] d t$.
\end{lemma}

{\noindent \sc Proof:}
Let us introduce coordinates $x^1,\ldots, x^{n}$ in a neighborhood
of $\gamma([0,1])$ in such a way that $\gamma(t)=(t,0,\ldots,0)$,
and let $P$ be a coordinate parallelepiped of sides $1, 2\epsilon,
\ldots, 2\epsilon$, and Lebesgue-coordinate volume
$V=(2\epsilon)^{n-1}$ around $\gamma$. By Rademacher's theorem $d f$
exists almost everywhere, that is, it exists on a measurable subset
$S$ of $P$ and $\int_P \chi_S=V$ where $\chi_S$ is the characteristic
function of $S$, and furthermore $d f=(\partial_1 f, \cdots,\partial_n f)$,
that is, wherever $df$ exists, the partial derivatives also exist
 (see
e.g.,~\cite{Heinonen05}).
 However, by Fubini's theorem $V=\int_P
\chi_S=\int_{-\epsilon}^{\epsilon} d x^n \cdots
\int_{-\epsilon}^{\epsilon} d x^2 \int_{0}^{1} d x^1 \chi_S$ which
proves that for almost all segments parallel to the $x^1$-axis we have $\int_{0}^{1} d x^1 \chi_S=1$, that is $d f$ exists almost
everywhere on almost every segment $\alpha(t)=(t,x^2,\cdots, x^{n})$
parallel to the image of $\gamma$. But clearly, wherever $d f$
exists on $\alpha$, $\partial_1 f=d f[e_1]=d f[\dot{\alpha}]$. Using
$\frac{d}{d t}(f\circ \alpha)=\partial_1 f$ we obtain $\frac{d}{d
t}(f\circ \alpha)=d f[\dot\alpha]$ almost everywhere on $[0,1]$.
Finally $f\circ \alpha$ is the composition of a $C^2$ function with
a locally Lipschitz function, thus locally Lipschitz and hence
absolutely continuous, from which the last identity follows.
\qedskip

%

We can now prove that the light cones can be opened preserving the
local anti-Lipschitz condition on the time function.

\begin{Theorem} \label{xhx}
 Let $(\mcM,g)$ be a stably causal spacetime with a
continuous metric $g$, and let $\nohattau$ be a time function on
$\mcM$. If  $\nohattau$ is locally Lipschitz
and locally $g$-anti-Lipschitz then the condition (*) of Theorem~\ref{T16IV12.1} holds, that is, there exists
a metric $\hg\succ g$  such that $\tau$ is locally
$\hat{g}$-anti-Lipschitz.
\end{Theorem}

{\noindent \sc Proof:}
Let us suppose that $\tau$ is locally Lipschitz and satisfies the
anti-Lipschitz condition on $g$-causal curves parametrised by
$h$-arc length, that is, for every compact set $K$ we can find
$C(K)>0$ such that
\begin{equation}
\nohattau(\gamma(s_2))- \nohattau(\gamma(s_1))\ge C(K) (s_2-s_1).
\end{equation}
Let $X$ be a $h$-normalized $g$-causal
vector, then taking the limit of this formula we find
\[
d \tau[X]\ge C(K)
\]
wherever $\tau$ is classically differentiable on $K$, hence almost
everywhere on $K$. We wish to prove that the inequality $d
\tau[Y]\ge C(K)/2$ holds wherever $d \tau$ exists on $K$, where $Y$
is a $h$-normalized $\hat{g}_K$-causal vector for $\hat{g}_K\succ g$
sufficiently close to $g$ on $K$. Unfortunately, we cannot use a
continuity argument because $d \tau$ exists only almost everywhere,
and is not necessarily continuous.

Suppose that $\tilde{g}_K\succ g$ is so close to $g$ that for any
$h$-normalized $\tilde{g}_K$-causal vector $Y$ we can find a
$h$-normalized ${g}$-causal vector $X$ such that $\Vert Y-X\Vert \le
C(K)/(2L)$, where $L$ is the Lipschitz constant of $\tau$ in $K$
(clearly, $\tilde{g}_K$ exists by a compactness argument). We have
\begin{align*}
\vert d \tau[X]-d \tau[Y]\vert &=\vert \lim_{t\to 0}[
\frac{\tau(p+Xt)-\tau(p)}{t}- \frac{\tau(p+Y t)-\tau(p)}{t}]\vert\\&\le
\lim_{t\to 0} \vert \frac{\tau(p+Xt)-\tau(p+Y t))}{t}\vert \le L \Vert
X-Y\Vert \le C(K)/2
\end{align*}
which implies $d \tau[Y]\ge C(K)/2$. Let $\hat{g}_K$ be such that $g\prec\hat{g}_K\prec\tilde{g}_K$, and let $t_1>t_0$.
Then given a $\hat{g}_K$-causal $h$-parametrised curve $\gamma(t)$, we
have by the previous lemma $\tau(\alpha(t_1))-\tau(\alpha(t_0))\ge
\frac{1}{2} C(K) (t_1-t_0)$ over a $h$-parametrised $\tilde{g}_K$-causal
curve $\alpha$ which we can take arbitrarily close to $\gamma$. Using the
continuity of $\tau$ we obtain $\tau(\gamma(t_1))-\tau(\gamma(t_0))\ge
\frac{1}{2} C(K) (t_1-t_0)$.

Let $K_i$ be a countable sequence of compact sets such that
$K_i\subset \textrm{Int} K_{i+1}$, $\cup_i K_i=\mcM$, and let
$\hat{g}_i$ be the metric just found for the compact set
$A_i=K_i\backslash \textrm{Int} K_{i-1}$. By making suitable
point-dependent convex combinations of $\hat g_i$ with $g$, we can
find $\hat{g}\succ g$ such that $\hat{g}\prec\hat{g}_i$ over every
$A_i$. Clearly, $\tau$  is locally $\hat{g}$-anti-Lipschitz, which
finishes the proof.
\qed

\medskip


The following
result, pointed out to us by A.~Fathi (private communication), turns out to be useful:

\begin{proposition}
 \label{P24XII12.1}
Let $(\mcM,g)$ be a spacetime admitting a time function $\tau$, then
for every $\epsilon>0$ there is a locally $g$-anti-Lipschitz time
function $\tilde{\tau}$ such that $\vert \tau-\tilde
\tau\vert<\epsilon$.
\end{proposition}

{\noindent \sc Proof:}
By Theorem \ref{pxr} there is a smooth time function with timelike
gradient $t: \mcM\to \mathbb{R}$.   Let
$\tau_\epsilon=\tau+\epsilon\tanh t$ so that $\vert \tau-
\tau_\epsilon \vert<\epsilon$. Since $t$ is locally
$g$-anti-Lipschitz (Proposition~\ref{bya}), so is $\tau_\epsilon$.
\qedskip

Given a locally Lipschitz time function $\nohattau$ we can use Proposition~\ref{P24XII12.1} to deform
$\nohattau$ to a time function which is both locally Lipschitz and anti-Lipschitz. By Theorems~\ref{xhx} and
Theorem~\ref{T16IV12.1} we conclude:

\begin{corollary}
 \label{C25XII12.11}
Let $(\mcM,g)$ be a stably causal spacetime with a continuous metric
$g$, and let $\nohattau$ be a time function on $\mcM$. If
$\nohattau$ is locally Lipschitz
 then
 for every $\epsilon$ there exists
a smooth $g$-time function $ {\tau}_\epsilon$, with $g$-timelike gradient,
such that $\vert {\tau}_\epsilon-\nohattau\vert <\epsilon$. If $\tau$ is Cauchy, then ${\tau}_\epsilon$ is also  Cauchy.
\qed
\end{corollary}

\begin{remark}
 \label{R25XII12.1}
Under the hypotheses of the corollary, if $\tau$ is further known to be locally anti-Lipschitz
then one likewise concludes that
for every function $\alpha \colon \mcM \to (0,+\infty)$ there exists
a smooth $g$-time function $ {\tau}_\alpha$, with $g$-timelike gradient,
such that $\vert  {\tau}_\alpha-\nohattau\vert <\alpha$.
\end{remark}

Recall that in~\cite{FathiSiconolfiTime} smooth time-functions are constructed on stably causal space-times,
by first constructing Lipschitz ones.
Corollary~\ref{C25XII12.11} gives
an alternative justification of the last step of the Fathi-Siconolfi
construction. We also note that the hypothesis that $\tau$ is Lipschitz is not necessary
for the conclusion of Corollary~\ref{C25XII12.11}, as any time functions can be approximated by
locally Lipschitz ones (A.~Fathi, private communication).

We can now apply  directly Theorem~\ref{T16IV12.1}  to the
continuously differentiable volume function with timelike gradient
obtained in Section~\ref{hgc}:

\begin{corollary}
 \label{c27XII12.1}
In a globally hyperbolic spacetime Geroch's Cauchy time function
$\tau=\ln ( \tau^-_\varphi/\tau^+_\varphi)$ is continuously
differentiable with timelike gradient for some $\varphi$, and
moreover, for such choice of $\varphi$ and for every function
$\alpha \colon \mcM \to (0,+\infty)$ there exists a smooth Cauchy time
function with timelike gradient, say $  {\tau}_\alpha$,  such that $\vert
\tau- {\tau}_\alpha\vert <\alpha$.
 \end{corollary}

{\noindent \sc Proof:}
The analysis of Section~\ref{hgc} shows that we can choose $\varphi$
so as to make $\tau^-_\varphi$ and $\tau^+_\varphi$ continuously
differentiable (and hence Lipschitz). Geroch's original argument
proves that $\tau$ is Cauchy. Theorem \ref{nju} proves that
$\tau^-_\varphi$ and $\tau^+_\varphi$ are locally $g$-anti-Lipschitz
and Theorem \ref{xhx} proves that for arbitrarily chosen positive
functions $\Delta \tau^\pm(x)$, there are smooth time functions with
timelike gradient $ {\tau}_\alpha^-$ and $- {\tau}_\alpha^+$, such that
$\vert\tau^-_\varphi-  {\tau}_\alpha^-\vert<\Delta \tau^-(x)$, and
$\vert\tau^+_\varphi-  {\tau}_\alpha^+\vert< \Delta \tau^+(x)$.

Let $x,y>0$ and $z=\ln(x/y)$ then $z(a',b')-z(a,b)=\int_{a;
y=b}^{a'} d x \frac{\partial z}{\partial x}+ \int_{b; x=a'}^{b'} d y
\frac{\partial z}{\partial y}$, and using the facts that
$\frac{\partial z}{\partial x}=\frac{1}{x}$, $-\frac{\partial
z}{\partial y}=\frac{1}{y}$ are decreasing, $\vert
z(a',b')-z(a,b)\vert \le \sup(\frac{1}{a'},\frac{1}{a}) \vert
a'-a\vert+ \sup(\frac{1}{b'},\frac{1}{b}) \vert b'-b\vert$.

Define $ {\tau}_\alpha= \ln (
{\tau}_\alpha^-/ {\tau}_\alpha^+)$, we then have at every point
$x\in \mcM$,
$$
 \vert \tau- {\tau}_\alpha\vert\le
 \frac{\Delta \tau^-}{\tau^-_\varphi-\Delta \tau^-}+ \frac{\Delta
 \tau^+}{\tau^+_\varphi-\Delta \tau^+}
 \;,
$$
wherever the denominators are positive. Choosing $\Delta \tau^\pm
\le \min{(\frac{1}{4}\alpha,\frac{1}{2})}\tau_\varphi^{\pm}$ we
obtain \[\vert \tau(x)- {\tau}_\alpha(x)\vert\le \min (\alpha(x),2)\]
(alternatively, this formula can be proved using the local Lipschitz
and local anti-Lipschitz character of $\tau^\pm_\varphi$ to show
that $\tau$ has the same properties). In particular, since the
right-hand side is bounded, $ {\tau}_\alpha$ is onto $\mathbb{R}$,
and since every level set of $ {\tau}_\alpha$ is contained in
$\tau^{-1}([a,b])$, that is, it stays between two Cauchy
hypersurfaces $\tau^{-1}(a)$, $\tau^{-1}(b)$, $ {\tau}_\alpha$ is
Cauchy.
\qed

\bigskip

{\sc \noindent Acknowledgements} PTC wishes to thank A.~Fathi, G.~Galloway and O.~M\"uller for useful discussions. He was supported in part by Narodowe Centrum Nauki under the grant DEC-2011/03/B/ST1/02625.
EM was partially supported  by GNFM of INDAM and by the Erwin Schr\"odinger Institute, Vienna.

\bigskip

%
%
%

\def\polhk#1{\setbox0=\hbox{#1}{\ooalign{\hidewidth
  \lower1.5ex\hbox{`}\hidewidth\crcr\unhbox0}}} \def\cprime{$'$}
  \def\cprime{$'$}
\providecommand{\bysame}{\leavevmode\hbox to3em{\hrulefill}\thinspace}
\providecommand{\MR}{\relax\ifhmode\unskip\space\fi MR }
\providecommand{\MRhref}[2]{%
  \href{http://www.ams.org/mathscinet-getitem?mr=#1}{#2}
}
\providecommand{\href}[2]{#2}

\end{document}